\documentclass[aps,prd,nofootinbib,twocolumn,longbibliography]{revtex4-1}
\usepackage[english]{babel}
\usepackage[autostyle, english = british]{csquotes}
\usepackage[CJKbookmarks, pdftex, bookmarksnumbered, bookmarksopen, colorlinks, citecolor=blue, linkcolor=blue]{hyperref}
\usepackage{amsmath,amssymb}
\usepackage{graphicx}
\usepackage{enumitem}
\usepackage{physics}
\setlist[enumerate]{topsep=0pt,parsep=-1mm,leftmargin=5mm,}
 \MakeOuterQuote{"}
\def\be{\begin{equation}}
\def\ee{\end{equation}}

\usepackage{xcolor}
\usepackage{xspace}
\usepackage{ifthen}

\newcommand{\showcomments}{true}

\newcommand{\andrea}[1]%
{\ifthenelse{\equal{\showcomments}{true}}%
{{\color{orange}{\small \textbf{Andrea says:} #1}}}{\xspace}}%

\newcommand{\marios}[1]%
{\ifthenelse{\equal{\showcomments}{true}}%
{{\color{blue}{\small \textbf{Marios:} #1}}}{\xspace}}%

\newcommand{\carlo}[1]%
{\ifthenelse{\equal{\showcomments}{true}}%
{{\color{purple}{\small \textbf{Carlo:} #1}}}{\xspace}}%

\begin{document}

\title{\large Gravity entanglement, quantum reference systems, degrees of freedom}

\author{Marios Christodoulou${}^{ab}$, Andrea Di Biagio${}^{a}$, Richard Howl${}^{cd}$, Carlo Rovelli${}^{efg}$}

\affiliation{\ }

\affiliation{${}^{a}$Institute for Quantum Optics and Quantum Information (IQOQI) Vienna, Austrian Academy of Sciences, Boltzmanngasse 3, A-1090 Vienna, Austria}
\affiliation{${}^{b}$Vienna Center for Quantum Science and Technology (VCQ), Faculty of Physics, University of Vienna, Boltzmanngasse 5, A-1090 Vienna, Austria}
\affiliation{${}^c$Quantum Group, Department of Computer Science, University of Oxford, Wolfson Building, Parks Road, Oxford, OX1 3QD, United Kingdom}
\affiliation{${}^d$QICI Quantum Information and Computation Initiative, Department of Computer Science, The University of Hong Kong, Pokfulam Road, Hong Kong}

\affiliation{${}^e$Aix-Marseille University, Universit\'e de Toulon, CPT-CNRS, F-13288 Marseille, France.}
\affiliation{${}^f$Department of Philosophy and the Rotman Institute of Philosophy, 1151 Richmond St.~N London  N6A5B7, Canada}
\affiliation{${}^g$Perimeter Institute, 31 Caroline Street N, Waterloo ON, N2L2Y5, Canada} 

\date{\today}
        
\begin{abstract} 

\noindent We discuss the physical interpretation of the gravity mediated entanglement effect. We show how to read it in terms of quantum reference systems. We pinpoint the single gravitational degree of freedom mediating the entanglement.  We clarify why the distinction between longitudinal and transverse degrees of freedom is irrelevant for the  interpretation of the results.  We discuss the relation between the LOCC theorem and the interpretation of the effect, its different relevance for, respectively, the quantum gravity and quantum information communities, and the reason for the excitement raised by the prospect of detection.  

\end{abstract}

\maketitle

\section{Introduction}

The suggestion that a  detection  of  Gravity Mediated Entanglement (GME) might be achieved in the lab \cite{bose2017spin,marletto2017gravitationallyinduced} has raised much interest  and a lively discussion on the precise significance of this detection (see \cite{huggett2022quantum} and full references there). Suggested  interpretations range from evidence for the quantum nature of geometry \cite{christodoulou2019possibility,christodoulou2022locally}, to agnosticism about the quantum or classical nature of the `true' gravitational degrees of freedom \cite{ anastopoulos2018comment,fragkos2022inference}.
Here we contribute to the discussion with several  observations. 

We consider the simplest version of GME: two particles of mass $m$ and positions $X$ and $Y$ are each set into a quantum superposition of two positions, say $X_\pm$ and $Y_\pm$, and kept in this state for a time $t$. Then the wave packets of both are recombined. In one of the four  branches the particles are kept at a short distance $d$ from each other. The gravitational potential energy $\delta E=-Gm^2/d$ in this branch, where $G$ is the Newton constant, gives rise to a difference in the total energy with respect to the other branches, that causes a relative shift in the phase of the branch
\begin{equation} 
\delta\phi = -\frac{\delta E}{\hbar}t=\frac{Gm^2t}{\hbar d}=\left(\frac{m}{m_\mathrm{Pl}}\right)^2\frac{ct}d,
\end{equation}
where $c$ is the speed of light and $m_\mathrm{Pl}$ the Planck mass.  The effect of the phase in one branch is that the state of the two particles can get entangled. (For complete  quantum field theoretical accounts of the effect, see for instance   \cite{christodoulou2022locally, Chen2022a}.) The entanglement can be detected by measuring degrees of freedom entangled with the position (for instance a spin) and checking the violation of the Bell inequalities.  This detection is evidence that the gravitational field is not classical, since two systems cannot be entangled by interacting via a third  {\em classical} system.  
Since we have learned from general relativity that the spacetime geometry measured by rods and clocks is determined by the gravitational field, a measurement of the GME effect is a detection of quantum properties of the physical geometry. 
This perspective has raised the lively interest of the quantum gravity community. 

A number of  cautionary remarks regarding the interpretation of the effect have been raised: 
\begin{enumerate} 
\item[(a)] The experiment is  in the non-relativistic regime: it cannot probe relativistic aspects of quantum gravity.  
\item[(b)] In this regime the notion of field can be dropped altogether and interactions can be described as instantaneous at a distance. Hence the experiment cannot probe aspects of the field nature of  gravity. 
\item[(c)] The experiment can be described using the field's longitudinal modes only. These are "enslaved" to matter, they are not part of the radiative modes generally  considered in  field quantization.    
  Hence the effect does not test  "genuine" quantum aspects of gravity  \cite{anastopoulos2018comment,anastopoulos2021gravitational}. 
\item[(d)] The impossibility of entangling degrees of freedom with a classical mediator can be shown in the context of the LOCC protocol \cite{horodecki2009quantum}, but the hypotheses of this protocol are not clearly satisfied in the experiment. 
\end{enumerate} 
In the following, we offer some considerations on the GME that shed light on these issues.  We discuss in particular the following questions
\begin{enumerate}
\item[i.] What exactly is the degree of freedom mediating entanglement, if any?
\item[ii.] What is the physical significance of the split between longitudinal and transverse modes? 
\item[iii.] What is the exact difference between this and past experiments  involving gravity and quantum matter?
\item[iv.] What does the experiment exactly provide evidence for, by itself or with additional assumptions? 
\end{enumerate} 
Some light on these questions is shed by viewing the experiment from the  `quantum reference frames' perspective  \cite{Rovelli1991,giacomini2019quantum}, as we do in the next section. 

\section{Quantum reference systems account of GME}

A fundamental tenet of general relativity is that the laws of physics are the same in all reference systems, where by "reference system" is meant here an arbitrary (smooth) choice of spacetime coordinates.  The relevance of reference systems for quantum mechanics and quantum gravity has long been noticed and has been recently exploited in a number of interesting situations \cite{AharonovY.Susskind1967,AharonovY.Susskind1984,AharonovY.Kaufherr1984,Rovelli1991,BartlettS.D.RudolphT.Spekkens2007,PalmerM.C.GirelliF.Bartlett2014,Castro-Ruiz,PereiraS.T.Angelo2015,giacomini2019quantum}. Since coordinates are arbitrary labels, nothing forbids us from using coordinates determined by the particles, rather than Cartesian coordinates determined by distances. Let us apply this logic to the GME setting. 

 We introduce general coordinates $x(X)$ such that $x(X)=0$ and $x(Y)=1$. In words, we can choose a coordinate system in which one particle has  coordinate  $x=0$ at all times and the other has  coordinate $x=1$ at all times.   We can change the spatial coordinates only,  without touching the time coordinate (defined a physical clock). 
 
 This can be done in all four branches where the particles are at respective  {\em Cartesian}  coordinate positions $(X_i, Y_j)$, where $i,j\in\{+,-\}$. Since the relation between the Cartesian and the new coordinates is different in each branch, the change of coordinates is different in the different branches, and the metric will then be different in the different branches.  Let $g^{(i,j)}_{ab}(x)$ be the metric in the $(i,j)$ branch,  in these  coordinates. 
 
 The distance between the two particles is then given by the standard general relativistic formula for the distance, which is the length of the shortest curve $\gamma$ on a fixed time surface. This definition is  diffeomorphism  invariant in this context. 
 Adapting the coordinates so that the curve $\gamma$ is along the $x$ direction, the distance is
\begin{equation} 
d_{(i,j)}(t)=\int_0^1  \sqrt{g^{(i,j)}_{xx}(x,t)} \ \dd x. 
\end{equation}
This formulation shows that the relevant degree of freedom for the experiment is a specific function of the gravitational field $g_{ab}(x,t)$: the line integral $d$, which has a different (time dependent) value $d_{(i,j)}(t)$ in each branch. In this coordinate system, the state of the system during the experiment can be written as 
\begin{equation} 
|\psi(t)\rangle \ =\ \  |X\!=\!0\rangle \otimes   |Y\!=\!1\rangle \otimes \sum_{{(i,j)}}|d_{(i,j)}(t)\rangle 
\end{equation}
where $|X=0\rangle$ and $|Y=1\rangle$ are the states of the particles and $|d\rangle$ is a (non normalised) quantum state of the  variable $d$ of the gravitational field.\footnote{Notice that both superposition and entanglement are frame-dependent, as nicely illustrated for example in \cite{giacomini2019quantum}.}

This drastic effect of the choice of reference frame brings forth the fact that in a gravitational context the physical degrees of freedom of the particles cannot be disentangled from those of the geometry. The relevant (relative) variable is the distance between the two masses, which is a non local function of the gravitational field.  Since the energy in each branch depends on this distance, the phase difference between the branches follows.  

Formally, the Hilbert space of the particles' positions plus the gravitational field is restricted by the diffeomorphism constraint, which forces the state to be independent from changes of coordinates. This allows us to consider only diffeomorphism invariant quantities, and $d$ is such a quantity: a relational quantity, that depends on both  field and particles. 

If the particles have a degree of freedom entangled with their distance, for instance a spin variable --or any other path dependent observable-- that takes the values $\pm$ in the respective $\pm$ branches, then this degree of freedom gets entangled with $d$:
\begin{equation} 
|\psi(t)\rangle \ =\  \sum_{i,j\in\{+,-\}}   |X\!=\!0, i\rangle \otimes   |Y\!=\!1, j\rangle \otimes |d_{(i,j)}(t)\rangle 
\end{equation}
Notice that in Cartesian coordinates the information about the distance is coded into the coordinate positions of the particle, and the spins get entangled with this position. In the reference frame we have chosen here, instead, the spins are directly entangled with the relevant variable, $d(t)$, whose dependence on the gravitational field is manifest. The coordinate position of the particles, which is pure gauge, is not affected by the dynamics and does not get entangled with anything.

Since $d(t)$ is time dependent during the experiment, the non-commuting observable that allows the mediating system to transfer entanglement are $d$ and its conjugate variable $p_d$.\footnote{In the Hamiltonian formulation of general relativity, the conjugate momentum to the spatial metric, hence also $d_p$, is related to the extrinsic curvature of spacetime \cite{Gravitation}.} When the particle is split into a superposition of distances, say with a Stern-Gerlach apparatus, the variable $d$ changes with time.  This can be modelled as an interaction between the particle's spin and $d$, which has the result of modifying the value of $d$. (The particles's position is gauge fixed in this reference frame.) If the initial spin of one particle is in the state $\ket{s_o}=c_+\ket{+}+c_-\ket{-}$ and the initial distance is $d_o$, the interaction's  evolution $U$ leads to
\begin{equation} 
U |s_o,d_o\rangle = \sum_i c_i\, |i,d_i\rangle. 
\end{equation}

Similarly, the evolution entangles the second particle's spin with $d$. Thus, the spins of the two particles will also get entangled \emph{through} $d$, giving rise to the four branches. 

The identification of $d$, namely the \emph{distance} between the particles, as the relevant entangling variable, emphasizes the fact that the GME effects can be interpreted as probing quantum properties of the spacetime  \emph{geometry}. Of course in physics geometry is operationally \emph{defined} by measurements performed with material instruments, and the astounding  empirical success of general relativity supports the fertility of Einstein's idea that this geometry is a manifestation of the gravitational field. Here we see all this in place. The GME apparatus can be seen as a devise indirectly measuring a distance, and the effect of its being in  superposition.

\section{The non-relativistic limit and the implications of the experiment}

In the previous section, we have employed general relativistic concepts and assumed that these can be used in this context.  In particular, we have assumed that the gravitational field and its function $d$ are dynamical variables.  The experiment, however, is  in the non-relativistic limit. As such, it can be described in non-relativistic terms.  In this language, each particle acts instantaneously on the other with a force. Such action-at-a distance describes well the world in this approximation.  In this picture, there is no mediator of the interaction and no degree of freedom associated to a mediator. 

This fact has raised a certain confusion in the discussion about the GME experiments.  More precisely, a positive outcome of the experiment is compatible with a world where {\em gravity has no special quantum properties and is an instantaneous interaction at a distance}.

Why then these experiments are considered relevant for quantum gravity in the quantum gravity community? The answer is that we know---{\em independently} from these specific experiments---that in the real world gravity is \emph{not} an instantaneous interaction at a distance, and---in fact---that there is no 
instantaneous interaction at a distance.  It is only by folding this additional knowledge in, that the experiment becomes informative for quantum gravity.  

In other words, while the experiment {\em alone} might not have strong implications, the experiment {\em in conjunction} with the fact that we have good reasons to believe, from a hundred years of experiments, that gravity is mediated by a field that propagates information at finite speeds \textit{does} have strong implications. If the information between the two particles is mediated by a field, and if the two particles get entangled, this does definitely count as evidence that the field cannot be a classical field. Therefore the  detection of the GME effect leaves us with an alternative:
either there is no such thing as the gravitational field, or this field cannot be a classical.

\section{Longitudinal and transverse components of the field}

The GME experiments are in the linear (weak field) regime. In this regime, the field can be usefully decomposed into a sum of components that can be treated differently. There are several ways to implement this decomposition.  To illustrate the fact, its implications, and also some common misunderstandings it generates, it is easier to discuss the electromagnetic version of the field-mediated entanglement effect, which is parallel but simpler \cite{chen2022quantum}.    In this case the masses have charges and the mediator is the electromagnetic field.  

A vector field $\vec A_\perp $ is called transverse if 
\begin{equation} 
\vec \nabla \cdot \vec A_\perp = 0,
\end{equation} 
while a vector field $A_\parallel$ is called longitudinal if it can be written as the gradient of a scalar function
\begin{equation} 
\vec A_\parallel = \vec \nabla f.
\end{equation} 
If $f$ is harmonic, namely it satisfies the Poisson equation $\Delta f=0$, the field $\vec A=\vec\nabla f$ is, according to these definitions, both longitudinal and transverse, but it is conventional to call it longitudinal.   If boundary conditions are fixed, the Poisson equation has a unique solution.   

In 1930, Enrico Fermi showed that the four dimensional Maxwell potential $A_\mu$ can be decomposed into two parts \cite{Fermi1930,Fermi1932}.  The first part, called longitudinal, is formed by the time component $A_0$ and the longitudinal part of the vector potential $\vec A_\parallel$; the second part, called radiative, is formed by the transverse (and non longitudinal) part of the vector potential $A_\perp$.  Fermi showed that the first part can be seen as giving rise to Coulomb interactions between particles, while the second part gives rise to the electromagnetic  radiation from moving charges. With this separation, one can use a non-relativistic form of the Hamiltonian for charged particles and fields (with self explanatory notation)
\begin{equation}  \label{EMHamiltonian}
H=\sum_n \frac{(\vec p_n-e_n\vec A_\perp(\vec x_n))^2}{2m_n}+\sum_{n>m}\frac{e_ne_m}{|\vec x_n-\vec x_m|}+H_\mathrm{rad}[A_\perp].
\end{equation}
where $H_\mathrm{rad}$ is the free field Hamiltonian that only depends on the transverse part of the field.  For a derivation, see for example \cite{cohen-tannoudji1989photons} chapter IB.

This decomposition and similar ones are useful and widely used in electromagnetism and in gravity, but it is important not to misinterpret them.  The decomposition into longitudinal and transverse parts (while gauge invariant) is not Lorentz invariant: it depends on the frame in which it is done.  Hence it cannot correspond to a physical distinction between different kinds of degrees of freedom of the theory.

In gravity the decomposition is even less well defined, because it can be done only in the vicinity of flat space.  

In general, there is no invariant sense in which a field splits into a radiative and a non radiative part: the split is always conventional. 

To understand why the  longitudinal/transverse split is unphysical, let us consider the following situation. A charge is kept at the origin until the time $t=0$. The field --say-- is the static Coulomb field centered on the charge.  During a finite time interval $\Delta t$ the charge is smoothly displaced from the origin to the new location $\vec x(t)\ne 0$, and then it is brought back to the origin. (That is $x(t) =0$ for all $t \notin [0, \Delta t]$ and $\vec x(t)\ne 0$ for some $t \in [0, \Delta t]$.) Consider the value of the electric field during the interval $\Delta t$ at a location $\vec y$ such that $|\vec y|\gg c\Delta t$.  Since the Maxwell equations are relativistic, the electric field $\vec E(\vec y)$ is obviously still the Coulomb field centered at the origin 
\begin{equation} 
\vec E(\vec y)=-\vec \nabla \frac{e}{|\vec y|}.
\end{equation}
The decomposition of the field into a longitudinal and a radiative part splits it as 
\begin{equation} 
\vec E(\vec y)=-\vec \nabla \frac{e}{|\vec y-\vec x(t)|}+\vec E_{\perp}(\vec y).
\end{equation}
The first term is the Coulomb field of the displaced source.  The second term, namely 
\begin{equation} 
\vec E_{\perp}(\vec y)\equiv -\vec \nabla \left( \frac{e}{|\vec y|}-\frac{e}{|\vec y-\vec x(t)|}\right),
\end{equation}
is a non-vanishing radiative part, which has magically appeared very far away from the charge, even if nothing is radiating there from the charge! Yet, it is necessary to account for the fact that the information about the change of location of the charge has not yet arrived at $\vec y$.

To interpret this decomposition as a decomposition into physically distinct degrees of freedom is obviously a physical nonsense: nothing happens at $\vec y$  when the charge at the origin is moved, because no information has had the time to get from the origin to $\vec y$.   
It is absurd to imagine that what happens in reality is that a physical part of the field instantaneously determined by the charges changes, and a radiative part magically appears to make this variation undetectable.

This example also shows that setting the transverse modes to zero means breaking the Maxwell equations even where there is no radiation  actually emitted by the charges. 
Also, notice that in the decomposition, not only the longitudinal part, but also the radiative part of the field has a badly non local dynamics. On this, and on analogous confusions permeating the gravitational waves literature, see also \cite{Ashtekar2017,Ashtekar2017a}.

The same argument applies in the gravity case.  Imagine that a fast neutron star arrives at extremely high speed from the sky and wipes away our sun.  For eight minutes, the Earth will continue to follow its curved orbit, being attracted by the sun, even if (in our Lorentz frame) the sun is not anymore at its place.   Since the Sun is eight light-minutes away, no physical information about its disappearance can affect the Earth's motion. This is  natural in field theory: the degrees of freedom of the field are local and they change at finite speed.  If we artificially separate the Newtonian from the radiative field, something absurd happens: the Newtonian field of the sun disappears instantaneously (in our Lorentz frame) as soon as the sun is wiped away by the neutron star, and a radiative mode appears, equally instantaneously, to compensate for its disappearance and  make it undetectable for eight minutes. 

The split into radiative and longitudinal modes is a computational trick that rearranges field components {\em globally} and does not respect the {\em local} nature of the physical degrees of freedom of the field. 

The moral is that Fermi's distinction between the Coulomb or Newton field components, and the radiative components of the field is a convenient mathematical trick that one can use in a frame and not a separation between physically distinct degrees of freedom. 
This conclusion is important because it has been argued that the component of the field involved in the Coulomb interaction is not a genuine field degree of freedom. This is incorrect. In the limit $c \rightarrow \infty$, the notion of field may be discarded altogether, but when $c$ is kept finite the appropriate account of the phenomena involves a propagating field and there is no physical distinction between genuine and not genuine parts of the field. 

\section{Can the radiative modes be quantum and the longitudinal modes be classical?}

Let us disregard the conclusion of the previous section and ignore the fact that the distinction between longitudinal and radiative modes of the field is not Lorentz invariant and not physically meaningful. Can we nevertheless say that the radiative parts of the field are affected by quantum theory, while the longitudinal parts are not? A moment of reflection shows that this would lead to contradictions. See also \cite{Belenchia2018,Belenchia2019a}.

In the GME experiments, the effect of the radiative part of the field is negligible and the field is always well approximated by its longitudinal part. If this remains classical,  the field would be entirely classical in the experiment.  But if it is classical, it cannot be in a state which is a quantum superposition of classical configurations. It must be in a single classical configuration, the same in all branches. Which one? Any such configuration would violate the Gauss law in at least one branch.  That is, for the effect to happen, it must be possible to have quantum superpositions of classical field configurations {\em for its longitudinal parts as well}.  In this sense the longitudinal part is equally quantum, and this is precisely what the GME experiments test.  

Let us see this in detail. In the Coulomb gauge, the longitudinal component $\vec A_\parallel$ of the  vector potential $\vec A$ is set to $0$. The longitudinal part of the electric field is  
\begin{equation}\label{Eparallel}
\vec E_\parallel= -\vec\nabla A_0
\end{equation}
and the transverse part is
\begin{equation}\label{Eperpendicular}
\vec E_\perp= -\partial_t \vec A.
\end{equation}
Thanks to the Gauss law, $A_0$ can be expressed completely in terms of variables of the matter sources. For a set of particles, for instance 
\begin{equation} 
A_0(x,t) = \sum_n\frac{q_n}{|\vec x_n(t)-\vec x|}.
\end{equation}
In the quantum theory in this gauge, the $\vec A(x)$ are operators on a Fock space, {\em but also the  $A_0(x)$ are operators}: they act on the Hilbert space of the matter sources. From \eqref{Eparallel} and \eqref{Eperpendicular} it then   follows that $\hat {\vec E}_\parallel$ acts on the  Hilbert space of matter while $\hat {\vec E}_\perp$ acts on the Fock space of the field. In this gauge (and in a fixed frame), we can say that the longitudinal field is "enslaved to matter". Nevertheless, this does not mean that it is classical: mathematically, $\hat {\vec E}_\parallel$ is very much a \textit{quantum field}: a family of operators (more precisely, operator distributions)  labeled by points in spacetime. Physically, precisely because matter is quantum, so is the longitudinal field: as  matter can be in quantum superposition, so can the longitudinal field. Concretely, the split between parallel and longitudinal fields is  unobservable: the electric field's observables are local quantities, sum of the two:
\begin{equation} 
\hat{\vec E}= \hat{\vec E}_\perp+\hat{\vec E}_\parallel.
\end{equation}
Therefore even disregarding the transverse modes, we can still  conclude that the field is in a quantum  superposition during the experiment. 

In gravity, as we have seen by changing reference frame, even the distinction between the positions of the particles and the  degrees of freedom of the field looses meaning, because general coordinate invariance (or, equivalently, because of presence of the diffeomorphism constraint) implies that the physical quantity relevant in the experiment mixes the two.

\section{The Quantum Information perspective}

A different source of confusion in the discussion on the physical relevance of the GME is rooted in the different interests and the different perspectives of the two main research communities concerned by this phenomenon.   In this regard, the history of the GME is peculiar.  Although motivated by quantum gravity, the idea of GME is rooted in the field of quantum information theory.  The idea of the experiment is in fact a beautiful case of cross fertilization between distant fields.   But the two communities work on the basis of different  assumptions and different scientific objectives.  

For the quantum gravity community, the momentous relevance of the GME experiments is, as mentioned, that that they can detect an effect of quantum superposition of geometry \cite{christodoulou2019possibility}.  The connection between GME detection and geometry superposition  gives for granted two basic facts: (a) the gravitational interaction is mediated by a local field, namely, a field that does not allow action at a distance, and (b) this field determines the spacetime geometry. These two ideas played a pivotal role in the development of our current theory of gravity (and modern physics in general) and are  generally taken for granted by the majority of quantum gravity researchers. 

The quantum information and quantum foundations community, on the other hand, is rooted in non-relativistic physics and is interested in particular in   theory-independent results: results that one can derive directly from the measurement, with minimal additional  theoretical inputs.  
In this logic, a number of assumptions considered obvious in the high energy community are not taken for granted. In this perspective, a relevant question is what does the detection of GME imply by itself, without additional knowledge about the nature of the gravitational interaction.  

For the vast majority of the important physical experiments, the result is  meaningful in the context of additional information and acquired knowledge about the world.  This is also a general point of philosophy of science of the last decades: there are no pure observations. An observation makes sense only within a rich context of assumptions, and is  interpreted as supporting, or questioning this context, as a whole. Without additional assumptions, it is rare to be able to extract useful information from a single observation. 

This fact has been explicit since the first proposal of GME.  The seminal paper \cite{bose2017spin} states:
\begin{quote}
    Our proposal relies on a simple assumption: the gravitational interaction between two masses is mediated by a gravitational field (in other words, it is not a direct interaction-at-a-distance).
\end{quote}
On the other hand, however, the same paper continues:
\begin{quote}
    Once we make this assumption, we use a central principle of quantum information theory: entanglement between two systems cannot be created by Local Operations and Classical Communication (LOCC).
\end{quote}
Let us see in more detail if this statement can be made precise. In the LOCC setup \cite{horodecki2009quantum}, one imagines two experimenters, each with full quantum control on a quantum system in their laboratory, that can coordinate their behaviour only by exchange of classical bits. Alice performs a generic operation on one system, communicates the classical outcome to Bob, who can use that information to pick what operation to perform on his system, send the outcome to Alice and so on. The relevant theorem states that entanglement is not generated, in this way.

Let us state the theorem more precisely. Call a \textit{local operation} on a bipartite system a \textit{separable quantum channel}, that is, a channel that takes the form $T_A\otimes T_B$ for some quantum channels $T_A$ and $T_B$. Here $T_A$ are Alice operations on one particle in her lab and $T_B$ are Bob operations on the other particle in his lab. A \textit{forward round of local operations and classical communication} is represented by a map of the form
\begin{equation}
    \rho\longmapsto\sum_i (\mathbb I\otimes T^i)\big[M_i(\rho)\otimes \mathbb I\big],
\end{equation}
where each $T^i$ is a quantum channel and the $\{M_i\}$ form an \textit{instrument}. Each $M_i$ is a completely positive map and  $\tr \sum_i M_i[\rho]=\tr \rho$. Here the label $i$ includes any signal or information transmitted by the field.   Bob's quantum channel depends on the field, and this in turn is affected by Alice's operation. We can define a \textit{backward round of local operations and classical communication} in total analogy. Two  rounds will look like
\begin{equation}
    \rho\longmapsto\sum_{ij} (T^j\otimes \mathbb I)\left[(\mathbb I\otimes \tilde M_j^i)\big[M_i(\rho)\otimes \mathbb I\big]\right],
\end{equation}
where each of the $T^j$ are channels, $\{M_i\}$ is an instrument, and for each $i$, $\{\tilde M^i_j\}$ is an instrument. We can keep defining $n$ rounds this way \cite{donald2002uniqueness}. 
The LOCC theorems show that the class of maps representing such exchanges is a subset of the \textit{separable operations}, those that can be represented with separable Kraus operators
\begin{equation}
    \rho\longmapsto\sum_i (A_i\otimes B_i)\rho(A_i\otimes B_i)^\dagger,
\end{equation}
where $\sum_i A_i^\dagger A_i \otimes B_i^\dagger B_i=\mathbb I$. These cannot entangle the quantum systems if they were originally in a tensor state. 

Notice that the notion of 
\textit{locality} employed in this theorem is only loosely connected with relativistic  locality. It refers simply to the split between the two subsystems and the fact Alice and Bob can each act only on one of the two. 
To apply the theorem to the GME effect, we have to identify the field with the classical communication channel, namely with any classical information exchanged between Alice and Bob's labs.  We can thus model the experiments as two quantum systems, namely the two particles, on which Alice and Bob intervene and that they can measure. 

Notice the subtle shift with respect to the usual quantum information setting.
What is meant here by (classical) information is any gravitational influence between the two labs. 
This information affects any operation or measurement Alice and Bob perform. In other words, the configuration of the gravitational field is in the index $i$ in the above formulas. Then the result follows: if any reciprocal gravitational influence of Alice's and Bob's operations is via the effect of gravity and this is a classical communication channel,  then the GME effect does not happen.

Of course by relaxing the definition of terms such as  \textit{local}, \textit{classical}, and \textit{communication},
entanglement can be generated in special situations.
In exotic quantum-classical hybrid models for instance  \cite{hall2018two}, it might in principle be possible for an evidently classical field to mediate entanglement between two quantum systems. 
Even in standard quantum theory one can turn a separable state into an entangled state\footnote{Prepare the quantum systems in either of two entangled states, and store the information of which state was prepared in a classical system. Then the state or the quantum systems alone is a separable state, but it is entangled when conditioning on the classical system.} by the exchange of a classical bit, \textit{if} the bit is correlated with the preparation of the quantum systems \cite{krisnanda2017revealing}.  Theory-independent claims about the experiment have to be understood in their own terms. The GME cannot be explained by a local classical mediator, under \emph{specific} definitions of these terms \emph{and} given extra assumptions (about the interactions, about superdeterminism etc). This is nothing new. The violation of the Bell inequalities show that the world is non-local, by a specific definition of non-local and assuming no-superdeterminism, the existence of a single world etc. Theory-independent results about experimental observations limit the space of possible theories. They can only “prove something” given extra assumptions. 

From the quantum gravity perspective, the full force of the GME experiments, as they have been conceived so far, is not in "proving" something: it is in confirming a general prediction that {\em in quantum gravity} comes about as a consequence of the fact that spacetime can be in superposition. 
In spite of its roots in quantum information, the main  interest of the experiment for quantum gravity research is not so much in the context of a theory independent framework.  The experiment is important because, given the current theoretical knowledge that we have about the world, GME is a direct consequence of quantum superposition of geometries.

\section{Difference from previous experiments}

A number of past experiments have involved quantum matter and geometry. For instance the COW experiment \cite{colella1975observation}, neutron bouncing \cite{jenke2011realization} and atomic fountain experiments can be interpreted relativistically as the measurement of the interference due to the different time dilation at different altitudes \cite{roura2022quantum,wootters2003why}.  None of these experiments, however, involve  actual quantum properties of geometry itself.  They are all compatible with a world where quantum matter moves in a classical geometry described by classical general relativity; for instance with the semiclassical theory defined by the Einstein equations coupled with the expectation value of the matter's energy momentum tensor $G=8
\pi\langle T\rangle$ \cite{wallace2021quantum}.

Not so the GME. In a world described by quantum matter on a classical geometry obeying general relativity, GME would not happen.  In this sense, detecting this effect  can be the first direct evidence of a quantum gravity phenomenon.

The fact that there is a regime in nature that is outside semiclassical gravity and yet gives an effect that survives in the $c\to\infty$ limit 
 is perhaps obvious a posteriori, but it is surprising and interesting.  
The GME effect is accounted for by a non-relativistic theory of the world where there is action at a distance and no quantum geometry, but this theory of the world is  falsified by a wide variety of {\em other} experiments.  Hence, {\em among the known viable theoretical alternatives}, the GME effect can only be interpreted as a manifestation of quantum geometry.

The true reason for the extreme interest raised by the prospect of GME detection is therefore subtle.  The GME effect is not predicted by semiclassical gravity, nor by most gravity induced collapse theories.  Hence it counts as evidence against these exotic alternatives.   But this is not the main reason of interest of the experiment.  After all, most scientists expect the effect to be real.  The reason of the interest is that as far as we can see today, the GME effect has only two possible explanations: (i) the world is genuinely non-relativistic or governed by something totally unknown, and this would  be astonishing, or (ii) the effect is a direct consequence of quantum geometry: seeing it could be our first direct glimpse into quantum gravity.

\begin{acknowledgments}
\textit{Acknowledgments--} This work was supported by the QISS (The Quantum Information Structure of Spacetime) John Templeton Foundation  Grant  No.61466 (qiss.fr). CR thanks Nick Huggett for interesting inputs and Flaminia Giacomini and Lin-Quing Chen for many discussions.  
\end{acknowledgments}

\bibliography{GMEQRF.bib}

\begin{thebibliography}{36}%
\makeatletter
\providecommand \@ifxundefined [1]{%
 \@ifx{#1\undefined}
}%
\providecommand \@ifnum [1]{%
 \ifnum #1\expandafter \@firstoftwo
 \else \expandafter \@secondoftwo
 \fi
}%
\providecommand \@ifx [1]{%
 \ifx #1\expandafter \@firstoftwo
 \else \expandafter \@secondoftwo
 \fi
}%
\providecommand \natexlab [1]{#1}%
\providecommand \enquote  [1]{``#1''}%
\providecommand \bibnamefont  [1]{#1}%
\providecommand \bibfnamefont [1]{#1}%
\providecommand \citenamefont [1]{#1}%
\providecommand \href@noop [0]{\@secondoftwo}%
\providecommand \href [0]{\begingroup \@sanitize@url \@href}%
\providecommand \@href[1]{\@@startlink{#1}\@@href}%
\providecommand \@@href[1]{\endgroup#1\@@endlink}%
\providecommand \@sanitize@url [0]{\catcode `\\12\catcode `\$12\catcode
  `\&12\catcode `\#12\catcode `\^12\catcode `\_12\catcode `\%12\relax}%
\providecommand \@@startlink[1]{}%
\providecommand \@@endlink[0]{}%
\providecommand \url  [0]{\begingroup\@sanitize@url \@url }%
\providecommand \@url [1]{\endgroup\@href {#1}{\urlprefix }}%
\providecommand \urlprefix  [0]{URL }%
\providecommand \Eprint [0]{\href }%
\providecommand \doibase [0]{http://dx.doi.org/}%
\providecommand \selectlanguage [0]{\@gobble}%
\providecommand \bibinfo  [0]{\@secondoftwo}%
\providecommand \bibfield  [0]{\@secondoftwo}%
\providecommand \translation [1]{[#1]}%
\providecommand \BibitemOpen [0]{}%
\providecommand \bibitemStop [0]{}%
\providecommand \bibitemNoStop [0]{.\EOS\space}%
\providecommand \EOS [0]{\spacefactor3000\relax}%
\providecommand \BibitemShut  [1]{\csname bibitem#1\endcsname}%
\let\auto@bib@innerbib\@empty
\bibitem [{\citenamefont {Bose}\ \emph {et~al.}(2017)\citenamefont {Bose},
  \citenamefont {Mazumdar}, \citenamefont {Morley}, \citenamefont {Ulbricht},
  \citenamefont {Toro{\v s}}, \citenamefont {Paternostro}, \citenamefont
  {Geraci}, \citenamefont {Barker}, \citenamefont {Kim},\ and\ \citenamefont
  {Milburn}}]{bose2017spin}%
  \BibitemOpen
  \bibfield  {author} {\bibinfo {author} {\bibfnamefont {Sougato}\ \bibnamefont
  {Bose}}, \bibinfo {author} {\bibfnamefont {Anupam}\ \bibnamefont {Mazumdar}},
  \bibinfo {author} {\bibfnamefont {Gavin~W.}\ \bibnamefont {Morley}}, \bibinfo
  {author} {\bibfnamefont {Hendrik}\ \bibnamefont {Ulbricht}}, \bibinfo
  {author} {\bibfnamefont {Marko}\ \bibnamefont {Toro{\v s}}}, \bibinfo
  {author} {\bibfnamefont {Mauro}\ \bibnamefont {Paternostro}}, \bibinfo
  {author} {\bibfnamefont {Andrew}\ \bibnamefont {Geraci}}, \bibinfo {author}
  {\bibfnamefont {Peter}\ \bibnamefont {Barker}}, \bibinfo {author}
  {\bibfnamefont {M.~S.}\ \bibnamefont {Kim}}, \ and\ \bibinfo {author}
  {\bibfnamefont {Gerard}\ \bibnamefont {Milburn}},\ }\bibfield  {title}
  {\enquote {\bibinfo {title} {A {{Spin Entanglement Witness}} for {{Quantum
  Gravity}}},}\ }\href {\doibase 10/gcsb22} {\bibfield  {journal} {\bibinfo
  {journal} {Physical Review Letters}\ }\textbf {\bibinfo {volume} {119}},\
  \bibinfo {pages} {240401} (\bibinfo {year} {2017})},\ \Eprint
  {http://arxiv.org/abs/1707.06050} {arXiv:1707.06050} \BibitemShut {NoStop}%
\bibitem [{\citenamefont {Marletto}\ and\ \citenamefont
  {Vedral}(2017)}]{marletto2017gravitationallyinduced}%
  \BibitemOpen
  \bibfield  {author} {\bibinfo {author} {\bibfnamefont {Chiara}\ \bibnamefont
  {Marletto}}\ and\ \bibinfo {author} {\bibfnamefont {Vlatko}\ \bibnamefont
  {Vedral}},\ }\bibfield  {title} {\enquote {\bibinfo {title}
  {Gravitationally-induced entanglement between two massive particles is
  sufficient evidence of quantum effects in gravity},}\ }\href {\doibase
  10/gcsjgn} {\bibfield  {journal} {\bibinfo  {journal} {Physical Review
  Letters}\ }\textbf {\bibinfo {volume} {119}},\ \bibinfo {pages} {240402}
  (\bibinfo {year} {2017})},\ \Eprint {http://arxiv.org/abs/1707.06036}
  {arXiv:1707.06036} \BibitemShut {NoStop}%
\bibitem [{\citenamefont {Huggett}\ \emph {et~al.}(2022)\citenamefont
  {Huggett}, \citenamefont {Linnemann},\ and\ \citenamefont
  {Schneider}}]{huggett2022quantum}%
  \BibitemOpen
  \bibfield  {author} {\bibinfo {author} {\bibfnamefont {Nick}\ \bibnamefont
  {Huggett}}, \bibinfo {author} {\bibfnamefont {Niels}\ \bibnamefont
  {Linnemann}}, \ and\ \bibinfo {author} {\bibfnamefont {Mike}\ \bibnamefont
  {Schneider}},\ }\href {http://arxiv.org/abs/2205.09013} {\enquote {\bibinfo
  {title} {Quantum {{Gravity}} in a {{Laboratory}}?}}\ } (\bibinfo {year}
  {2022}),\ \Eprint {http://arxiv.org/abs/2205.09013} {arXiv:2205.09013}
  \BibitemShut {NoStop}%
\bibitem [{\citenamefont {Christodoulou}\ and\ \citenamefont
  {Rovelli}(2019)}]{christodoulou2019possibility}%
  \BibitemOpen
  \bibfield  {author} {\bibinfo {author} {\bibfnamefont {Marios}\ \bibnamefont
  {Christodoulou}}\ and\ \bibinfo {author} {\bibfnamefont {Carlo}\ \bibnamefont
  {Rovelli}},\ }\bibfield  {title} {\enquote {\bibinfo {title} {On the
  possibility of laboratory evidence for quantum superposition of
  geometries},}\ }\href {\doibase 10/gj6ssc} {\bibfield  {journal} {\bibinfo
  {journal} {Physics Letters B}\ }\textbf {\bibinfo {volume} {792}},\ \bibinfo
  {pages} {64--68} (\bibinfo {year} {2019})},\ \Eprint
  {http://arxiv.org/abs/1808.05842} {arXiv:1808.05842} \BibitemShut {NoStop}%
\bibitem [{\citenamefont {Christodoulou}\ \emph {et~al.}(2022)\citenamefont
  {Christodoulou}, \citenamefont {Di~Biagio}, \citenamefont {Aspelmeyer},
  \citenamefont {Brukner}, \citenamefont {Rovelli},\ and\ \citenamefont
  {Howl}}]{christodoulou2022locally}%
  \BibitemOpen
  \bibfield  {author} {\bibinfo {author} {\bibfnamefont {Marios}\ \bibnamefont
  {Christodoulou}}, \bibinfo {author} {\bibfnamefont {Andrea}\ \bibnamefont
  {Di~Biagio}}, \bibinfo {author} {\bibfnamefont {Markus}\ \bibnamefont
  {Aspelmeyer}}, \bibinfo {author} {\bibfnamefont {{\v C}aslav}\ \bibnamefont
  {Brukner}}, \bibinfo {author} {\bibfnamefont {Carlo}\ \bibnamefont
  {Rovelli}}, \ and\ \bibinfo {author} {\bibfnamefont {Richard}\ \bibnamefont
  {Howl}},\ }\href {http://arxiv.org/abs/2202.03368} {\enquote {\bibinfo
  {title} {Locally mediated entanglement through gravity from first
  principles},}\ } (\bibinfo {year} {2022}),\ \Eprint
  {http://arxiv.org/abs/2202.03368} {arXiv:2202.03368} \BibitemShut {NoStop}%
\bibitem [{\citenamefont {Anastopoulos}\ and\ \citenamefont
  {Hu}(2018)}]{anastopoulos2018comment}%
  \BibitemOpen
  \bibfield  {author} {\bibinfo {author} {\bibfnamefont {C.}~\bibnamefont
  {Anastopoulos}}\ and\ \bibinfo {author} {\bibfnamefont {Bei-Lok}\
  \bibnamefont {Hu}},\ }\href {http://arxiv.org/abs/1804.11315} {\enquote
  {\bibinfo {title} {Comment on ``{{A Spin Entanglement Witness}} for {{Quantum
  Gravity}}'' and on ``{{Gravitationally Induced Entanglement}} between {{Two
  Massive Particles}} is {{Sufficient Evidence}} of {{Quantum Effects}} in
  {{Gravity}}''},}\ } (\bibinfo {year} {2018}),\ \Eprint
  {http://arxiv.org/abs/1804.11315} {arXiv:1804.11315} \BibitemShut {NoStop}%
\bibitem [{\citenamefont {Fragkos}\ \emph {et~al.}(2022)\citenamefont
  {Fragkos}, \citenamefont {Kopp},\ and\ \citenamefont
  {Pikovski}}]{fragkos2022inference}%
  \BibitemOpen
  \bibfield  {author} {\bibinfo {author} {\bibfnamefont {Vasileios}\
  \bibnamefont {Fragkos}}, \bibinfo {author} {\bibfnamefont {Michael}\
  \bibnamefont {Kopp}}, \ and\ \bibinfo {author} {\bibfnamefont {Igor}\
  \bibnamefont {Pikovski}},\ }\href {http://arxiv.org/abs/2206.00558} {\enquote
  {\bibinfo {title} {On inference of quantization from gravitationally induced
  entanglement},}\ } (\bibinfo {year} {2022}),\ \Eprint
  {http://arxiv.org/abs/2206.00558} {arXiv:2206.00558} \BibitemShut {NoStop}%
\bibitem [{\citenamefont {Chen}\ \emph
  {et~al.}(2022{\natexlab{a}})\citenamefont {Chen}, \citenamefont {Giacomini},\
  and\ \citenamefont {Rovelli}}]{Chen2022a}%
  \BibitemOpen
  \bibfield  {author} {\bibinfo {author} {\bibfnamefont {Lin-Qing}\
  \bibnamefont {Chen}}, \bibinfo {author} {\bibfnamefont {Flaminia}\
  \bibnamefont {Giacomini}}, \ and\ \bibinfo {author} {\bibfnamefont {Carlo}\
  \bibnamefont {Rovelli}},\ }\bibfield  {title} {\enquote {\bibinfo {title}
  {{Quantum States of Fields for Quantum Split Sources}},}\ }\href@noop {}
  {\bibfield  {journal} {\bibinfo  {journal} {to appear}\ } (\bibinfo {year}
  {2022}{\natexlab{a}})}\BibitemShut {NoStop}%
\bibitem [{\citenamefont {Anastopoulos}\ \emph {et~al.}(2021)\citenamefont
  {Anastopoulos}, \citenamefont {Lagouvardos},\ and\ \citenamefont
  {Savvidou}}]{anastopoulos2021gravitational}%
  \BibitemOpen
  \bibfield  {author} {\bibinfo {author} {\bibfnamefont {Charis}\ \bibnamefont
  {Anastopoulos}}, \bibinfo {author} {\bibfnamefont {Mihalis}\ \bibnamefont
  {Lagouvardos}}, \ and\ \bibinfo {author} {\bibfnamefont {Konstantina}\
  \bibnamefont {Savvidou}},\ }\bibfield  {title} {\enquote {\bibinfo {title}
  {Gravitational effects in macroscopic quantum systems: A first-principles
  analysis},}\ }\href {\doibase 10/gpbc7q} {\bibfield  {journal} {\bibinfo
  {journal} {Classical and Quantum Gravity}\ }\textbf {\bibinfo {volume}
  {38}},\ \bibinfo {pages} {155012} (\bibinfo {year} {2021})},\ \Eprint
  {http://arxiv.org/abs/2103.08044} {arXiv:2103.08044} \BibitemShut {NoStop}%
\bibitem [{\citenamefont {Horodecki}\ \emph {et~al.}(2009)\citenamefont
  {Horodecki}, \citenamefont {Horodecki}, \citenamefont {Horodecki},\ and\
  \citenamefont {Horodecki}}]{horodecki2009quantum}%
  \BibitemOpen
  \bibfield  {author} {\bibinfo {author} {\bibfnamefont {Ryszard}\ \bibnamefont
  {Horodecki}}, \bibinfo {author} {\bibfnamefont {Pawel}\ \bibnamefont
  {Horodecki}}, \bibinfo {author} {\bibfnamefont {Michal}\ \bibnamefont
  {Horodecki}}, \ and\ \bibinfo {author} {\bibfnamefont {Karol}\ \bibnamefont
  {Horodecki}},\ }\bibfield  {title} {\enquote {\bibinfo {title} {Quantum
  entanglement},}\ }\href {\doibase 10/d2vqp8} {\bibfield  {journal} {\bibinfo
  {journal} {Reviews of Modern Physics}\ }\textbf {\bibinfo {volume} {81}},\
  \bibinfo {pages} {865--942} (\bibinfo {year} {2009})},\ \Eprint
  {http://arxiv.org/abs/quant-ph/0702225} {arXiv:quant-ph/0702225} \BibitemShut
  {NoStop}%
\bibitem [{\citenamefont {Rovelli}(1991)}]{Rovelli1991}%
  \BibitemOpen
  \bibfield  {author} {\bibinfo {author} {\bibfnamefont {C.}~\bibnamefont
  {Rovelli}},\ }\bibfield  {title} {\enquote {\bibinfo {title} {{Quantum
  reference systems}},}\ }\href {\doibase 10.1088/0264-9381/8/2/012} {\bibfield
   {journal} {\bibinfo  {journal} {Classical and Quantum Gravity}\ }\textbf
  {\bibinfo {volume} {8}},\ \bibinfo {pages} {317--331} (\bibinfo {year}
  {1991})}\BibitemShut {NoStop}%
\bibitem [{\citenamefont {Giacomini}\ \emph {et~al.}(2019)\citenamefont
  {Giacomini}, \citenamefont {{Castro-Ruiz}},\ and\ \citenamefont
  {Brukner}}]{giacomini2019quantum}%
  \BibitemOpen
  \bibfield  {author} {\bibinfo {author} {\bibfnamefont {Flaminia}\
  \bibnamefont {Giacomini}}, \bibinfo {author} {\bibfnamefont {Esteban}\
  \bibnamefont {{Castro-Ruiz}}}, \ and\ \bibinfo {author} {\bibfnamefont {{\v
  C}aslav}\ \bibnamefont {Brukner}},\ }\bibfield  {title} {\enquote {\bibinfo
  {title} {Quantum mechanics and the covariance of physical laws in quantum
  reference frames},}\ }\href {\doibase 10/gfv3xs} {\bibfield  {journal}
  {\bibinfo  {journal} {Nature Communications}\ }\textbf {\bibinfo {volume}
  {10}},\ \bibinfo {pages} {494} (\bibinfo {year} {2019})},\ \Eprint
  {http://arxiv.org/abs/1712.07207} {arXiv:1712.07207} \BibitemShut {NoStop}%
\bibitem [{\citenamefont {{Aharonov, Y. ,
  Susskind}}(1967)}]{AharonovY.Susskind1967}%
  \BibitemOpen
  \bibfield  {author} {\bibinfo {author} {\bibfnamefont {L.}~\bibnamefont
  {{Aharonov, Y. , Susskind}}},\ }\bibfield  {title} {\enquote {\bibinfo
  {title} {{Charge Superselection Rule}},}\ }\href {\doibase
  10.1103/PhysRev.155.1428} {\bibfield  {journal} {\bibinfo  {journal} {Phys.
  Rev.}\ }\textbf {\bibinfo {volume} {155}},\ \bibinfo {pages} {1428} (\bibinfo
  {year} {1967})}\BibitemShut {NoStop}%
\bibitem [{\citenamefont {{Aharonov, Y.,
  Susskind}}(1967)}]{AharonovY.Susskind1984}%
  \BibitemOpen
  \bibfield  {author} {\bibinfo {author} {\bibfnamefont {L.}~\bibnamefont
  {{Aharonov, Y., Susskind}}},\ }\bibfield  {title} {\enquote {\bibinfo {title}
  {{Observability of the Sign Change of Spinors under 2$\pi$ Rotations}},}\
  }\href {\doibase 10.1103/PhysRev.158.1237} {\bibfield  {journal} {\bibinfo
  {journal} {Phys. Rev. D}\ }\textbf {\bibinfo {volume} {158}},\ \bibinfo
  {pages} {1237} (\bibinfo {year} {1967})}\BibitemShut {NoStop}%
\bibitem [{\citenamefont {{Aharonov, Y.
  Kaufherr}}(1984)}]{AharonovY.Kaufherr1984}%
  \BibitemOpen
  \bibfield  {author} {\bibinfo {author} {\bibfnamefont {T.}~\bibnamefont
  {{Aharonov, Y. Kaufherr}}},\ }\bibfield  {title} {\enquote {\bibinfo {title}
  {{Quantum frames of reference}},}\ }\href {\doibase 10.1103/PhysRevD.30.368}
  {\bibfield  {journal} {\bibinfo  {journal} {Phys. Rev. D}\ }\textbf {\bibinfo
  {volume} {30}},\ \bibinfo {pages} {368} (\bibinfo {year} {1984})}\BibitemShut
  {NoStop}%
\bibitem [{\citenamefont {{Bartlett, S. D., Rudolph, T.,
  Spekkens}}(2007)}]{BartlettS.D.RudolphT.Spekkens2007}%
  \BibitemOpen
  \bibfield  {author} {\bibinfo {author} {\bibfnamefont {R.~W.}\ \bibnamefont
  {{Bartlett, S. D., Rudolph, T., Spekkens}}},\ }\bibfield  {title} {\enquote
  {\bibinfo {title} {{Reference frames, superselection rules, and quantum
  information}},}\ }\href {\doibase 10.1103/RevModPhys.79.555} {\bibfield
  {journal} {\bibinfo  {journal} {Rev Mod Phys}\ }\textbf {\bibinfo {volume}
  {79}},\ \bibinfo {pages} {555} (\bibinfo {year} {2007})},\ \Eprint
  {http://arxiv.org/abs/quant-ph/0610030} {arXiv:quant-ph/0610030} \BibitemShut
  {NoStop}%
\bibitem [{\citenamefont {{Palmer, M. C., Girelli, F.,
  Bartlett}}(2014)}]{PalmerM.C.GirelliF.Bartlett2014}%
  \BibitemOpen
  \bibfield  {author} {\bibinfo {author} {\bibfnamefont {S.~D.}\ \bibnamefont
  {{Palmer, M. C., Girelli, F., Bartlett}}},\ }\bibfield  {title} {\enquote
  {\bibinfo {title} {{Changing quantum reference frames}},}\ }\href {\doibase
  10.1103/PhysRevA.89.052121} {\bibfield  {journal} {\bibinfo  {journal} {Phys
  rev A}\ }\textbf {\bibinfo {volume} {89}},\ \bibinfo {pages} {052121}
  (\bibinfo {year} {2014})},\ \Eprint {http://arxiv.org/abs/1307.6597}
  {arXiv:1307.6597} \BibitemShut {NoStop}%
\bibitem [{\citenamefont {Castro-Ruiz}\ \emph {et~al.}(2020)\citenamefont
  {Castro-Ruiz}, \citenamefont {Giacomini}, \citenamefont {Belenchia},\ and\
  \citenamefont {Brukner}}]{Castro-Ruiz}%
  \BibitemOpen
  \bibfield  {author} {\bibinfo {author} {\bibfnamefont {E.}~\bibnamefont
  {Castro-Ruiz}}, \bibinfo {author} {\bibfnamefont {F.}~\bibnamefont
  {Giacomini}}, \bibinfo {author} {\bibfnamefont {A.}~\bibnamefont
  {Belenchia}}, \ and\ \bibinfo {author} {\bibfnamefont {{\v{C}}.}~\bibnamefont
  {Brukner}},\ }\bibfield  {title} {\enquote {\bibinfo {title} {{Quantum clocks
  and the temporal localisability of events in the presence of gravitating
  quantum systems}},}\ }\href@noop {} {\bibfield  {journal} {\bibinfo
  {journal} {Nature communication}\ }\textbf {\bibinfo {volume} {11}},\
  \bibinfo {pages} {1--2} (\bibinfo {year} {2020})}\BibitemShut {NoStop}%
\bibitem [{\citenamefont {{Pereira, S. T.
  Angelo}}(2015)}]{PereiraS.T.Angelo2015}%
  \BibitemOpen
  \bibfield  {author} {\bibinfo {author} {\bibfnamefont {R.~M.}\ \bibnamefont
  {{Pereira, S. T. Angelo}}},\ }\bibfield  {title} {\enquote {\bibinfo {title}
  {{Galilei covariance and Einstein's equivalence principle in quantum
  reference frames}},}\ }\href {\doibase 10.1103/PhysRevA.91.022107} {\bibfield
   {journal} {\bibinfo  {journal} {Phys rev A}\ }\textbf {\bibinfo {volume}
  {91}},\ \bibinfo {pages} {022107} (\bibinfo {year} {2015})},\ \Eprint
  {http://arxiv.org/abs/1404.2908} {arXiv:1404.2908} \BibitemShut {NoStop}%
\bibitem [{\citenamefont {Misner}\ \emph {et~al.}(1973)\citenamefont {Misner},
  \citenamefont {Thorne},\ and\ \citenamefont {Wheeler}}]{Gravitation}%
  \BibitemOpen
  \bibfield  {author} {\bibinfo {author} {\bibfnamefont {Charles~W}\
  \bibnamefont {Misner}}, \bibinfo {author} {\bibfnamefont {Kip}\ \bibnamefont
  {Thorne}}, \ and\ \bibinfo {author} {\bibfnamefont {John~A}\ \bibnamefont
  {Wheeler}},\ }\href@noop {} {\emph {\bibinfo {title} {{Gravitation (Physics
  Series)}}}}\ (\bibinfo  {publisher} {W. H. Freeman},\ \bibinfo {year}
  {1973})\BibitemShut {NoStop}%
\bibitem [{\citenamefont {Chen}\ \emph
  {et~al.}(2022{\natexlab{b}})\citenamefont {Chen}, \citenamefont {Giacomini},\
  and\ \citenamefont {Rovelli}}]{chen2022quantum}%
  \BibitemOpen
  \bibfield  {author} {\bibinfo {author} {\bibfnamefont {Lin-Qing}\
  \bibnamefont {Chen}}, \bibinfo {author} {\bibfnamefont {Flaminia}\
  \bibnamefont {Giacomini}}, \ and\ \bibinfo {author} {\bibfnamefont {Carlo}\
  \bibnamefont {Rovelli}},\ }\href@noop {} {\enquote {\bibinfo {title} {Quantum
  fields of quantum split sources},}\ } (\bibinfo {year}
  {2022}{\natexlab{b}})\BibitemShut {NoStop}%
\bibitem [{\citenamefont {Fermi}(1930)}]{Fermi1930}%
  \BibitemOpen
  \bibfield  {author} {\bibinfo {author} {\bibfnamefont {Enrico}\ \bibnamefont
  {Fermi}},\ }\bibfield  {title} {\enquote {\bibinfo {title} {{Sopra
  l'elettrodinamica quantstica}},}\ }\href@noop {} {\bibfield  {journal}
  {\bibinfo  {journal} {Rendiconti Lincei}\ }\textbf {\bibinfo {volume} {12}},\
  \bibinfo {pages} {431--435} (\bibinfo {year} {1930})}\BibitemShut {NoStop}%
\bibitem [{\citenamefont {Fermi}(1932)}]{Fermi1932}%
  \BibitemOpen
  \bibfield  {author} {\bibinfo {author} {\bibfnamefont {Enrico}\ \bibnamefont
  {Fermi}},\ }\bibfield  {title} {\enquote {\bibinfo {title} {{Quantum Theory
  of Radiation}},}\ }\href@noop {} {\bibfield  {journal} {\bibinfo  {journal}
  {Review of Modern Physics}\ }\textbf {\bibinfo {volume} {4}},\ \bibinfo
  {pages} {87--123} (\bibinfo {year} {1932})}\BibitemShut {NoStop}%
\bibitem [{\citenamefont {{Cohen-Tannoudji}}\ \emph {et~al.}(1989)\citenamefont
  {{Cohen-Tannoudji}}, \citenamefont {{Dupont-Roc}},\ and\ \citenamefont
  {Grynberg}}]{cohen-tannoudji1989photons}%
  \BibitemOpen
  \bibfield  {author} {\bibinfo {author} {\bibfnamefont {Claude}\ \bibnamefont
  {{Cohen-Tannoudji}}}, \bibinfo {author} {\bibfnamefont {Jacques}\
  \bibnamefont {{Dupont-Roc}}}, \ and\ \bibinfo {author} {\bibfnamefont
  {Gilbert}\ \bibnamefont {Grynberg}},\ }\href@noop {} {\emph {\bibinfo {title}
  {{Photons and atoms: introduction to quantum electrodynamics}}}}\ (\bibinfo
  {publisher} {{Wiley}},\ \bibinfo {address} {{New York}},\ \bibinfo {year}
  {1989})\BibitemShut {NoStop}%
\bibitem [{\citenamefont {Ashtekar}\ and\ \citenamefont
  {Bonga}(2017{\natexlab{a}})}]{Ashtekar2017}%
  \BibitemOpen
  \bibfield  {author} {\bibinfo {author} {\bibfnamefont {Abhay}\ \bibnamefont
  {Ashtekar}}\ and\ \bibinfo {author} {\bibfnamefont {B{\'{e}}atrice}\
  \bibnamefont {Bonga}},\ }\bibfield  {title} {\enquote {\bibinfo {title} {{On
  the ambiguity in the notion of transverse traceless modes of gravitational
  waves}},}\ }\href {\doibase 10.1007/s10714-017-2290-z} {\bibfield  {journal}
  {\bibinfo  {journal} {General Relativity and Gravitation}\ }\textbf {\bibinfo
  {volume} {49}},\ \bibinfo {pages} {1--43} (\bibinfo {year}
  {2017}{\natexlab{a}})},\ \Eprint {http://arxiv.org/abs/1707.09914}
  {arXiv:1707.09914} \BibitemShut {NoStop}%
\bibitem [{\citenamefont {Ashtekar}\ and\ \citenamefont
  {Bonga}(2017{\natexlab{b}})}]{Ashtekar2017a}%
  \BibitemOpen
  \bibfield  {author} {\bibinfo {author} {\bibfnamefont {Abhay}\ \bibnamefont
  {Ashtekar}}\ and\ \bibinfo {author} {\bibfnamefont {B{\'{e}}atrice}\
  \bibnamefont {Bonga}},\ }\bibfield  {title} {\enquote {\bibinfo {title} {{On
  a basic conceptual confusion in gravitational radiation theory}},}\ }\href
  {\doibase 10.1088/1361-6382/aa88e2} {\bibfield  {journal} {\bibinfo
  {journal} {Classical and Quantum Gravity}\ }\textbf {\bibinfo {volume}
  {34}},\ \bibinfo {pages} {20LT01} (\bibinfo {year} {2017}{\natexlab{b}})},\
  \Eprint {http://arxiv.org/abs/1707.07729} {arXiv:1707.07729} \BibitemShut
  {NoStop}%
\bibitem [{\citenamefont {Belenchia}\ \emph {et~al.}(2018)\citenamefont
  {Belenchia}, \citenamefont {Wald}, \citenamefont {Giacomini}, \citenamefont
  {Castro-Ruiz}, \citenamefont {Brukner},\ and\ \citenamefont
  {Aspelmeyer}}]{Belenchia2018}%
  \BibitemOpen
  \bibfield  {author} {\bibinfo {author} {\bibfnamefont {Alessio}\ \bibnamefont
  {Belenchia}}, \bibinfo {author} {\bibfnamefont {Robert~M.}\ \bibnamefont
  {Wald}}, \bibinfo {author} {\bibfnamefont {Flaminia}\ \bibnamefont
  {Giacomini}}, \bibinfo {author} {\bibfnamefont {Esteban}\ \bibnamefont
  {Castro-Ruiz}}, \bibinfo {author} {\bibfnamefont {{\v{C}}aslav}\ \bibnamefont
  {Brukner}}, \ and\ \bibinfo {author} {\bibfnamefont {Markus}\ \bibnamefont
  {Aspelmeyer}},\ }\bibfield  {title} {\enquote {\bibinfo {title} {{Quantum
  Superposition of Massive Objects and the Quantization of Gravity}},}\ }\href
  {\doibase 10.1103/PhysRevD.98.126009} {\bibfield  {journal} {\bibinfo
  {journal} {Physical Review D}\ }\textbf {\bibinfo {volume} {98}} (\bibinfo
  {year} {2018}),\ 10.1103/PhysRevD.98.126009},\ \Eprint
  {http://arxiv.org/abs/1807.07015v2} {arXiv:1807.07015v2} \BibitemShut
  {NoStop}%
\bibitem [{\citenamefont {Belenchia}\ \emph {et~al.}(2019)\citenamefont
  {Belenchia}, \citenamefont {Wald}, \citenamefont {Giacomini}, \citenamefont
  {Castro-Ruiz}, \citenamefont {Brukner},\ and\ \citenamefont
  {Aspelmeyer}}]{Belenchia2019a}%
  \BibitemOpen
  \bibfield  {author} {\bibinfo {author} {\bibfnamefont {Alessio}\ \bibnamefont
  {Belenchia}}, \bibinfo {author} {\bibfnamefont {Robert~M.}\ \bibnamefont
  {Wald}}, \bibinfo {author} {\bibfnamefont {Flaminia}\ \bibnamefont
  {Giacomini}}, \bibinfo {author} {\bibfnamefont {Esteban}\ \bibnamefont
  {Castro-Ruiz}}, \bibinfo {author} {\bibfnamefont {Aaslav}\ \bibnamefont
  {Brukner}}, \ and\ \bibinfo {author} {\bibfnamefont {Markus}\ \bibnamefont
  {Aspelmeyer}},\ }\bibfield  {title} {\enquote {\bibinfo {title} {{Information
  content of the gravitational field of a quantum superposition}},}\ }\href
  {\doibase 10.1142/S0218271819430016} {\bibfield  {journal} {\bibinfo
  {journal} {International Journal of Modern Physics D}\ }\textbf {\bibinfo
  {volume} {28}} (\bibinfo {year} {2019}),\ 10.1142/S0218271819430016},\
  \Eprint {http://arxiv.org/abs/1905.04496} {arXiv:1905.04496} \BibitemShut
  {NoStop}%
\bibitem [{\citenamefont {Donald}\ \emph {et~al.}(2002)\citenamefont {Donald},
  \citenamefont {Horodecki},\ and\ \citenamefont
  {Rudolph}}]{donald2002uniqueness}%
  \BibitemOpen
  \bibfield  {author} {\bibinfo {author} {\bibfnamefont {Matthew~J.}\
  \bibnamefont {Donald}}, \bibinfo {author} {\bibfnamefont {Michal}\
  \bibnamefont {Horodecki}}, \ and\ \bibinfo {author} {\bibfnamefont {Oliver}\
  \bibnamefont {Rudolph}},\ }\bibfield  {title} {\enquote {\bibinfo {title}
  {The {{Uniqueness Theorem}} for {{Entanglement Measures}}},}\ }\href
  {\doibase 10.1063/1.1495917} {\bibfield  {journal} {\bibinfo  {journal}
  {Journal of Mathematical Physics}\ }\textbf {\bibinfo {volume} {43}},\
  \bibinfo {pages} {4252--4272} (\bibinfo {year} {2002})},\ \Eprint
  {http://arxiv.org/abs/quant-ph/0105017} {arXiv:quant-ph/0105017} \BibitemShut
  {NoStop}%
\bibitem [{\citenamefont {Hall}\ and\ \citenamefont
  {Reginatto}(2018)}]{hall2018two}%
  \BibitemOpen
  \bibfield  {author} {\bibinfo {author} {\bibfnamefont {Michael J.~W.}\
  \bibnamefont {Hall}}\ and\ \bibinfo {author} {\bibfnamefont {Marcel}\
  \bibnamefont {Reginatto}},\ }\bibfield  {title} {\enquote {\bibinfo {title}
  {On two recent proposals for witnessing nonclassical gravity},}\ }\href
  {\doibase 10.1088/1751-8121/aaa734} {\bibfield  {journal} {\bibinfo
  {journal} {Journal of Physics A: Mathematical and Theoretical}\ }\textbf
  {\bibinfo {volume} {51}},\ \bibinfo {pages} {085303} (\bibinfo {year}
  {2018})},\ \Eprint {http://arxiv.org/abs/1707.07974} {arXiv:1707.07974}
  \BibitemShut {NoStop}%
\bibitem [{\citenamefont {Krisnanda}\ \emph {et~al.}(2017)\citenamefont
  {Krisnanda}, \citenamefont {Zuppardo}, \citenamefont {Paternostro},\ and\
  \citenamefont {Paterek}}]{krisnanda2017revealing}%
  \BibitemOpen
  \bibfield  {author} {\bibinfo {author} {\bibfnamefont {Tanjung}\ \bibnamefont
  {Krisnanda}}, \bibinfo {author} {\bibfnamefont {Margherita}\ \bibnamefont
  {Zuppardo}}, \bibinfo {author} {\bibfnamefont {Mauro}\ \bibnamefont
  {Paternostro}}, \ and\ \bibinfo {author} {\bibfnamefont {Tomasz}\
  \bibnamefont {Paterek}},\ }\bibfield  {title} {\enquote {\bibinfo {title}
  {Revealing non-classicality of inaccessible objects},}\ }\href {\doibase
  10/gcvfp4} {\bibfield  {journal} {\bibinfo  {journal} {Physical Review
  Letters}\ }\textbf {\bibinfo {volume} {119}},\ \bibinfo {pages} {120402}
  (\bibinfo {year} {2017})},\ \Eprint {http://arxiv.org/abs/1607.01140}
  {arXiv:1607.01140} \BibitemShut {NoStop}%
\bibitem [{\citenamefont {Colella}\ \emph {et~al.}(1975)\citenamefont
  {Colella}, \citenamefont {Overhauser},\ and\ \citenamefont
  {Werner}}]{colella1975observation}%
  \BibitemOpen
  \bibfield  {author} {\bibinfo {author} {\bibfnamefont {R.}~\bibnamefont
  {Colella}}, \bibinfo {author} {\bibfnamefont {A.~W.}\ \bibnamefont
  {Overhauser}}, \ and\ \bibinfo {author} {\bibfnamefont {S.~A.}\ \bibnamefont
  {Werner}},\ }\bibfield  {title} {\enquote {\bibinfo {title} {Observation of
  {{Gravitationally Induced Quantum Interference}}},}\ }\href {\doibase
  10/dktp8g} {\bibfield  {journal} {\bibinfo  {journal} {Physical Review
  Letters}\ }\textbf {\bibinfo {volume} {34}},\ \bibinfo {pages} {1472--1474}
  (\bibinfo {year} {1975})}\BibitemShut {NoStop}%
\bibitem [{\citenamefont {Jenke}\ \emph {et~al.}(2011)\citenamefont {Jenke},
  \citenamefont {Geltenbort}, \citenamefont {Lemmel},\ and\ \citenamefont
  {Abele}}]{jenke2011realization}%
  \BibitemOpen
  \bibfield  {author} {\bibinfo {author} {\bibfnamefont {Tobias}\ \bibnamefont
  {Jenke}}, \bibinfo {author} {\bibfnamefont {Peter}\ \bibnamefont
  {Geltenbort}}, \bibinfo {author} {\bibfnamefont {Hartmut}\ \bibnamefont
  {Lemmel}}, \ and\ \bibinfo {author} {\bibfnamefont {Hartmut}\ \bibnamefont
  {Abele}},\ }\bibfield  {title} {\enquote {\bibinfo {title} {Realization of a
  gravity-resonance-spectroscopy technique},}\ }\href {\doibase
  10.1038/nphys1970} {\bibfield  {journal} {\bibinfo  {journal} {Nature
  Physics}\ }\textbf {\bibinfo {volume} {7}},\ \bibinfo {pages} {468--472}
  (\bibinfo {year} {2011})}\BibitemShut {NoStop}%
\bibitem [{\citenamefont {Roura}(2022)}]{roura2022quantum}%
  \BibitemOpen
  \bibfield  {author} {\bibinfo {author} {\bibfnamefont {Albert}\ \bibnamefont
  {Roura}},\ }\bibfield  {title} {\enquote {\bibinfo {title} {Quantum probe of
  space-time curvature},}\ }\href {\doibase 10.1126/science.abm6854} {\bibfield
   {journal} {\bibinfo  {journal} {Science}\ }\textbf {\bibinfo {volume}
  {375}},\ \bibinfo {pages} {142--143} (\bibinfo {year} {2022})}\BibitemShut
  {NoStop}%
\bibitem [{\citenamefont {Wootters}(2003)}]{wootters2003why}%
  \BibitemOpen
  \bibfield  {author} {\bibinfo {author} {\bibfnamefont {William~K.}\
  \bibnamefont {Wootters}},\ }\bibfield  {title} {\enquote {\bibinfo {title}
  {Why {{Things Fall}}},}\ }\href {\doibase 10/bgt3fd} {\bibfield  {journal}
  {\bibinfo  {journal} {Foundations of Physics}\ }\textbf {\bibinfo {volume}
  {33}},\ \bibinfo {pages} {1549--1557} (\bibinfo {year} {2003})}\BibitemShut
  {NoStop}%
\bibitem [{\citenamefont {Wallace}(2021)}]{wallace2021quantum}%
  \BibitemOpen
  \bibfield  {author} {\bibinfo {author} {\bibfnamefont {David}\ \bibnamefont
  {Wallace}},\ }\href {http://arxiv.org/abs/2112.12235} {\enquote {\bibinfo
  {title} {Quantum {{Gravity}} at {{Low Energies}}},}\ } (\bibinfo {year}
  {2021}),\ \Eprint {http://arxiv.org/abs/2112.12235} {arXiv:2112.12235}
  \BibitemShut {NoStop}%
\end{thebibliography}%

\end{document}